\newcommand{\diracslash}[1]{#1\llap{/\kern2pt}}

\newcommand{\be}{\begin{equation}}
\newcommand{\ee}{\end{equation}}
\newcommand{\bea}{\begin{eqnarray}}
\newcommand{\eea}{\end{eqnarray}}
\newcommand{\ba}[1]{\begin{array}{#1}}
\newcommand{\ea}{\end{array}}

\documentclass[twocolumn,amssymb,nobibnotes,nofootinbib,aps,pra,showpacs]{revtex4}
\usepackage{graphicx}
\begin{document}
\setlength{\topmargin}{-0.05in}

\title{Bragg scattering of light in a strongly interacting trapped  Fermi gas of atoms}

\author{Bimalendu Deb}
 \affiliation{ Physical
Research Laboratory, Navrangpura, Ahmedabad 380 009, India}

\date{\today}

\def\be{\begin{equation}}
\def\ee{\end{equation}}
\def\bearr{\begin{eqnarray}}
\def\eearr{\end{eqnarray}}
\def\zbf#1{{\bf {#1}}}
\def\bfm#1{\mbox{\boldmath $#1$}}
\def\hf{\frac{1}{2}}

\begin{abstract}
We study Bragg scattering of laser light by trapped Fermi atoms
having two hyperfine spin components in the unitarity-limited
strongly interacting regime at zero temperature.  We calculate the
dynamic structure function of the superfluid trapped Fermi gas in
the unitarity limit.  Model calculation using local density
approximation shows that, the superfluid pairing gap in the
unitarity limit is detectable from the measurements of dynamic
structure function by Bragg spectroscopy, while in the
weak-coupling BCS limit, the gap eludes such spectroscopic
detection.

\end{abstract}


\maketitle

\section{introduction}

Achieving Bardeen-Cooper-Schrieffer (BCS) type phase transition in
trapped neutral fermionic atoms is the primary goal in the current
experimental research with ultracold atomic Fermi gases. Recently,
two groups \cite{duke,innsbruke} have measured collective
oscillations in strongly interacting trapped Fermi atoms. The
results of their measurements suggest the occurrence of fermionic
superfluidity in the atoms according to a theoretical prediction
\cite{stringari}. Condensation of ``fermionic atom pairs" has been
reported earlier \cite{colorado}. The superfluid pairing is
believed to occur near the crossover \cite{randeria,crossover}
between the predicted BCS state of fermionic atoms and the
Bose-Einstein  condensation of bosonic dimer molecules formed from
Fermi atoms due to a magnetic field Feshbach resonance. Several
groups have experimentally produced Bose-Einstein condensates
(BEC) \cite{molecules} of molecules formed from degenerate Fermi
atoms in the vicinity of the Feshbach resonance. The atoms become
strongly interacting at and near the Feshbach resonance. Several
other recent experiments \cite{kmo,expt1,expt2,expt3} with Fermi
atoms have demonstrated the strong-coupling behavior of the atoms.
Therefore, many-body effects become important in describing the
physics near the predicted BCS-BEC crossover in Fermi atoms.
Recently, a number of theoretical investigations \cite{theory}
have revealed many intriguing aspects of many-body effects in the
crossover regime. Despite the  recent experimental indications of
the occurrence of Cooper-like pairing near the crossover, the
energy gap for such pairing has not been so far measured due to
the lack of efficient detection methods.

There are several methods available for measuring pairing gap in
electronic superconductors, but any of them is hardly applicable
to the trapped neutral Fermi atoms. Recently, several theoretical
proposals \cite{zoller,huletp} have been made for detecting atomic
Cooper pairs by laser light.  There has been  a suggestion
\cite{zoller} to use resonant light for exciting one of the
Cooper-paired atoms into an excited electronic level, and thus to
make an interface between normal and superfluid state in analogy
with the well known tunnelling experiment in superconductors. Our
purpose here is to examine Bragg scattering of lasers as a method
for measuring the paring gap of a superfluid trapped Fermi gas of
atoms. Bragg spectroscopy has been already used for measuring
structure function of an atomic BEC \cite{bragg}. Unlike in a BEC,
the analysis of Bragg scattering in a superfluid trapped atomic
Fermi gas is complicated due mainly to the Pauli blocking and the
paring gap.

In this paper, we study Bragg scattering of light in a
two-component trapped atomic Fermi gas at zero temperature.  Bragg
scattering in a Fermi gas of atoms has an analogy with Raman
scattering in electronic superconductors.  We therefore, develop
our theoretical treatment following the theory of Raman scattering
in superconductors \cite{abrikosov}. The relevant physical
quantity is the dynamic structure function which is the Fourier
transform of density-density correlation function in time
\cite{nozpines}. It is a measure of the spectrum of density
fluctuation and proportional to the differential scattering cross
section per unit energy. In the case of superconductors,  it is
well known that unless the energy transferred from a probe to the
superconductor exceeds $2\Delta$ ($\Delta$ being the pairing gap),
the dynamic structure function is zero.  At $2\Delta$, it shows a
sharp discontinuity. In the case of superfluid trapped Fermi gas,
$\Delta$ has a spatial distribution varying from a maximum at the
trap center to a vanishingly  small value at the edge of the trap.
We find that the dynamic structure function of a superfluid
trapped Fermi gas in the unitarity limit has a prominent shift
compared to  that of a normal or superfluid trapped Fermi gas in
the BCS limit. It also exhibits a discontinuity at $
2\Delta({\mathbf 0})$ where $\Delta({\mathbf 0})$ refers to the
energy gap at the trap center. As the energy transfer decreases
below $2\Delta({\mathbf 0})$, the dynamic structure function falls
off with reducing slope,  and as the energy transfer goes to zero,
the slope vanishes. In contrast, in the case  of a normal trapped
gas or a superfluid trapped gas in the BCS limit, the dynamic
structure function increases almost linearly with the energy
transfer in the low energy regime.

The paper is organized as follows. In the following section, we
discuss in some detail the  physical scenario for Bragg scattering
in two-component trapped Fermi atoms in the unitarity-limited
strongly interacting regime. In Sec.III, we discuss the
theoretical treatment for calculating dynamic structure function
of a uniform superfluid Fermi gas. We then generalize this
treatment for a superfluid trapped Fermi gas in Sec.IV. The
results are discussed in Sec.V. We conclude in Sec.VI.

\section{Trapped Fermi gas in the unitarity regime}

We consider a harmonic  trap with potential $V_{ho}(r,z) = (1/2)m
( \omega_{\perp} r_{\perp}^2 + \omega_z z^2)$, where $
\omega_{\perp}$ and $\omega_z$ denote the radial and axial
frequency, respectively, of the trap. The harmonic oscillator is
characterized by the radial (axial) length scale $a_{\perp (z)}=
\sqrt{\hbar/(m\omega_{\perp (z)}}$. One can define a geometric
mean frequency of the harmonic oscillator by $\omega_{ho} =
(\omega_{\perp}^2\omega_z)^{1/3}$ and a corresponding geometric
mean length scale of the oscillator by $a_{ho} =
\sqrt{\hbar/(m\omega_{ho})}$. In our treatment, we resort to
Thomas-Fermi local density approximation (LDA) \cite{houbiers}
which is particularly applicable when the local Fermi energy is
larger than the average level spacing of the trap and the
coherence length of the fermion-pair is shorter than the average
trap size. Under this approximation,  the state of the system is
governed by \bearr \epsilon_F({\mathbf r})+ V_{ho}({\mathbf r}) +
U({\mathbf r}) = \mu \label{nr} \eearr where $\epsilon_F ({\mathbf
r}) = \hbar^2 k_F({\mathbf r})^2 /(2m)$ is the local Fermi energy,
$k_F({\mathbf r})$ denotes the local Fermi momentum which is
related to the local number density by $n({\mathbf r}) =
k_F({\mathbf r})^3/(6\pi^2)$. Here $U$ represents the mean-field
interaction energy and $\mu$ is the chemical potential fixed  by
normalization condition. At low energy, the mean-field interaction
energy depends on the two-body s-wave scattering amplitude
$f_0(k)= -a_s/(1+ia_sk)$, where $a_s$ represents
energy-independent s-wave scattering length and $k$ denotes the
relative wave number of two colliding particles. In the dilute gas
limit ($|a_s|k <\!<1$), $U$ becomes proportional to $a_s$ in the
form $U({\mathbf r}) = \frac{4\pi\hbar^2 a_s}{2m} n({\mathbf r})$.
 In the unitarity limit $|a_s|k \rightarrow \infty$,
the scattering amplitude $f_0 \sim i/k$ and hence  $U$ becomes
independent of $a_s$. It then follows from a simple dimensional
analysis that in this limit, $U$ should be proportional to the
Fermi energy: $U({\mathbf r}) = \beta \epsilon_F({\mathbf r})$
where $\beta$ is the constant of proportionality. Under LDA,  the
density profile of a trapped Fermi  gas is given by
 \bearr
n({\mathbf r}) &=& n({\mathbf 0}) \left [1
-\frac{r_{\perp}^2}{R_{\perp}^2} - \frac{r_{z}^2}{R_z^2} \right
]^{3/2} \label{nr} \eearr where \bearr n({\mathbf 0}) =
\frac{1}{6\pi^2\hbar^3}\left(\frac{2m\mu }{1+\beta}\right)^{3/2}
\eearr is the density of the atoms at the trap center. Here
$R_{\perp (z)}^2 = 2\mu/(m\omega_{\perp (z)}^2)$ being the
radial(axial) Thomas-Fermi radius. The normalization condition on
Eq. (\ref{nr}) gives an expression for  \bearr
 \mu = (1+\beta)^{1/2} (6N_{\sigma})^{1/3}\hbar\omega_0
\eearr where $N_{\sigma}$  is the total number of atoms in the
hyperfine spin $\sigma$. The Fermi momentum $k_F = [3\pi^2
n({\mathbf 0})]^{1/3}$ is then given by
\begin{eqnarray}
k_F =  \frac{1}{\hbar}\sqrt{\frac{2m\mu}{1+\beta}}
=(1+\beta)^{-1/4}k_F^0
\end{eqnarray}
where \bearr k_F^0 = \frac{(48N_{\sigma})^{1/6}}{a_{ho}} \eearr is
the Fermi momentum of the noninteracting trapped gas. For an
attractive interaction ($-1<\beta<0$ ) in the unitarity limit,
Fermi momentum would be larger than that of noninteracting gas by
a factor of $(1+\beta)^{-1/4}$. Accordingly, the Fermi energy
would be larger than the Fermi energy $\epsilon_F^0 = \hbar
\omega_F^0$ of noninteracting gas  by a factor of
$(1+\beta)^{-1/2}$.

To illustrate the main idea, we specifically consider trapped
$^6$Li Fermi atoms in their two lowest hyperfine spin  states
$\mid g_1 \rangle = \mid 2{\rm S}_{1/2}, F = 1/2, m_{F} = 1/2
\rangle$ and $\mid g_2 \rangle = \mid 2{\rm S}_{1/2}, F=1/2, m_F =
-1/2 \rangle $. For s-wave pairing to occur, the atom number
difference $\delta N $ of the two components should be restricted
by $\frac{\delta N}{N} \le T_c/\epsilon_F$ where $T_c$ is the
critical temperature for superfluid transition and $\epsilon_F$ is
the Fermi energy at the trap center.  Unequal densities of the two
components result in interior gap superfluidity
\cite{wilczek,mishra} which naturally arises in QCD matter. For
simplicity, we consider the case $N_{1/2} = N_{-1/2}$ which is the
optimum condition for s-wave Cooper pairing. An applied magnetic
field tuned near the Feshbach resonance ($\sim 822$ Gauss) results
in  strong inter-component s-wave interaction. At such high
magnetic fields, the splitting between the two ground hyperfine
states is $\sim 75$ MHz \cite{zwierlein}, while the corresponding
splitting between the excited states $ \mid e_1 \rangle = \mid
2{\rm P}_{3/2}, F=3/2, m_{F} = -1/2 \rangle $ and $\mid e_2\rangle
= \mid 2{\rm P}_{3/2}, F = 3/2, m_{F} = -3/2 \rangle $ is $\sim
994$ MHz \cite{thomas}. Taking advantage of these Zeeman splits,
it is possible to scatter atoms selectively of one hyperfine spin
component by making use of a pair of circularly polarized laser
beams-both having either $\sigma_{+}$ or $\sigma_{-}$
polarization.

In Bragg scattering, two laser beams (Bragg pulses) with a small
frequency difference  are applied to the trapped gas.  The
magnitude of this momentum transfer is $q \simeq 2 k_L
\sin(\theta/2) $, where $\theta$ is the angle between the two
beams and $k_L$ is the momentum of a laser photon. Let  both the
laser beams be $\sigma_{-}$ polarized and tuned near the
transition $\mid g_2\rangle \rightarrow \mid e_2\rangle$. Then the
transition between the states $\mid g_1 \rangle $ and $\mid e_2
\rangle $ would be forbidden while the transition $\mid g_1
\rangle \rightarrow \mid e_1\rangle $ will be suppressed due to
the large detuning $\sim 900$ MHz. This leads to a situation where
the Bragg-scattered atoms remain in the same initial internal
state $\mid e_2\rangle$. Similarly, atoms in state $\mid g_1
\rangle$ only would undergo Bragg scattering when two $
\sigma_{+}$ polarized lasers are tuned near the transition $\mid
g_1 \rangle \rightarrow \mid 2{\rm P}_{3/2}, F = 3/2, m_F =
3/2\rangle$. Thus, we infer that in the presence of a high
magnetic field,  it is possible to scatter atoms selectively of
either spin components only by using circularly polarized Bragg
lasers.

For simplicity, let us first consider Bragg scattering in a
uniform Fermi gas. Later, we will analyze the nonuniform case. We
assume that both the laser beams are $\sigma_-$ polarized and
tuned  near the transition $\mid g_2\rangle \rightarrow \mid e_2
\rangle $.  The detuning of both the lasers from the atomic
resonance should be much greater than the single- and two-photon
line widths in order to avoid heating of the system. Under such
conditions, for a uniform gas, the laser-atom interaction
Hamiltonian can be expressed as
 \bearr
 H_{af} \simeq \hbar \Omega e^{-i\delta
 t} \sum_{{\mathbf k}}
  \hat{c}_{\downarrow}^{\dagger}({\mathbf k }+ {\mathbf q}) \hat{c}_{\downarrow}({\mathbf
  k})+
 {\mathrm H.c.}
  \eearr
where $\hat{c}_{\downarrow}({\mathbf k})$ represents the atomic
field operator with momentum ${\mathbf k}$, the subscript
$\downarrow$ ($\uparrow$) refers to the state $\mid g_2\rangle$
($\mid g_1\rangle$) and $\Omega$ is the two-photon Rabi frequency.
Here $\delta = \omega_1 - \omega_2$ is the detuning between the
two lasers.  One can identify the operator
$\hat{\rho}_{\downarrow}^{\dagger}({\mathbf q}) = \sum_{{\mathbf
k}} \hat{c}_{\downarrow}^{\dagger}({\mathbf{k + q}})
\hat{c}_{\downarrow}({\mathbf k})$ as the Fourier transform of the
density operator $\hat{\rho}_{\downarrow}({\mathbf r}) =
\hat{\Psi}_{\downarrow}^{\dagger}({\mathbf
r})\hat{\Psi}_{\downarrow}({\mathbf r})$ where
$\hat{\Psi}_{\downarrow}^{\dagger}({\mathbf r})$ represents the
field operator in the real space. The spectrum of the scattered
atoms would be proportional to the rate of transition probability
which, according to Fermi's Golden rule, is given by \bearr \kappa
&=& 2\pi\hbar \Omega^2 \sum_f | \langle f \mid
\hat{\rho}_{\downarrow}^{\dagger}({\mathbf q}) \mid 0 \rangle|^2
\delta(\hbar\delta - \epsilon_f + \epsilon_0) \nonumber \\
&=& 2\pi\hbar \Omega^2 S({\mathbf q},\omega) \eearr where
$|0\rangle$ represents the many-body ground state and sum runs
over all the final states $|f\rangle$ which can be coupled to the
ground state by the operator $\hat{\rho}_{\downarrow}^{\dagger}$.

\section{dynamic structure function of a uniform superfluid}

 Using Bogliubov transformations \bearr
\hat{c}_{\downarrow}({\mathbf k}) = u_k
\hat{\gamma}_{\downarrow}({\mathbf k}) + v_k^*
\hat{\gamma}_{\uparrow}^{\dagger}(-{\mathbf k}) \\
\nonumber \\
\hat{c}_{\uparrow}({\mathbf k}) = u_k
\hat{\gamma}_{\uparrow}({\mathbf k}) - v_k^*
\hat{\gamma}_{\downarrow}^{\dagger}(-{\mathbf k}) \eearr one can
reexpress the interaction Hamiltonian in terms of quasiparticle
operators $\hat{\gamma}_{\sigma}^{\dagger}({\mathbf k})$ and
$\hat{\gamma}_{\sigma}({\mathbf k})$. Here the normalization
condition is $|u_k|^2 + |v_k|^2 =1$ with \bearr |v_k|^2 = (1/2)(1
- \xi_k/E_k) \eearr  and \bearr E_k = \sqrt{\xi_k^2 + \Delta_k^2},
\eearr where $\xi_k = \hbar^2k^2/(2m) - \mu$ and $\Delta_k$ is the
pairing gap. The chemical potential $\mu$ and the gap $\Delta_k$
can be obtained by solving the regularized gap equation \bearr
\frac{m}{4\pi\hbar^2a_s} = \frac{1}{V} \sum_{{\mathbf k}} \left
(\frac{1}{2\epsilon_k} - \frac{1}{2E_k)}\right ) \label{gap}
\eearr along with the equation \bearr n= \frac{1}{6\pi^2} k_F^3 =
\frac{1}{V} \sum_{{\mathbf k}} \left (1- \frac{\xi_k}{E_k}\right
). \label{num} \eearr of the density of single component.  Here
$V$ is the volume of the system. The analytical solutions of these
two coupled equations have been already obtained in the Ref
\cite{analyt} for the entire range of the parameter $a_sk_F$
starting from weak interaction ($a_sk_F\rightarrow \pm 0$) to the
unitarity limit ($a_sk_F\rightarrow \pm \infty$). In the unitarity
limit, the mean-field interaction becomes independent of the
scattering length and the Fermi system exhibits universal
behavior. In this limit, the only energy scale available to the
system is the Fermi energy.   The mean-field energy  and the gap
then become proportional to the Fermi energy. Solutions of the
above two equations in the unitarity limit yield $\Delta \simeq
1.16\mu$, $\mu = (1+\beta)\epsilon_F$, where $\beta = -0.41$
\cite{analyt,pitaevskii} is a constant. Calculations using quantum
monte carlo simulation \cite{phand} yields $\beta = -0.56$ and
$\Delta \simeq 0.49 \epsilon_F$, while calculations using
Galitskii and lowest order constraint variation (LOCV)
approximations \cite{heiselberg} give $\beta = -0.67$ and $\beta =
-0.43$, respectively. The constant $\beta$ was experimentally
introduced and measured to be -0.26 in Ref \cite{kmo}, while it
was experimentally found to be between -0.4 to -0.3 in Ref
\cite{expt2}. A recent experiment \cite{barten} has obtained
$\beta = -0.68$.

One can notice that the BCS ground state $|0\rangle$, which is
annihilated by the quasiparticle operators, will have nonzero
matrix element of $\hat{\rho}_{\mathbf q}$ only when the final
state $|f\rangle$ is a state of two quasiparticles with mutually
opposite spins and their momenta differing by ${\mathbf q}$. The
dynamic structure function thus takes the form \bearr S({\mathbf
q},\delta) &=&
 \sum_{{\rm f}}|u_{k'}v_k|^2\delta(\hbar \delta - E_{k'} - E_{k}) \nonumber \\
&=& \frac{V}{4(2\pi)^3}\int d^3{\mathbf k} \frac{(E_{k'} +
\xi_{k'})(E_k
- \xi_k)}{E_{k'}E_k} \nonumber \\
&\times& \delta(\hbar \delta - E_{k'} - E_{k}) \label{dsfu} \eearr
where $k' = |{\mathbf k} + {\mathbf q}|$ is the wave vector of a
scattered atom.  Note that the usual BCS coherence factor
$m({\mathbf k}, {\mathbf k}')=u_{k'}v_{k} + u_{k}v_{k'}$
\cite{schreiffer,abrikosov} which appears in the description of
electronic superconductors has changed. This is due to the fact
that the polarization-selective dipole transitions in Fermi atoms
in the presence of strong magnetic field as  discussed earlier
lead to the transfer of momentum and energy to either partner (of
hyperfine spin $\uparrow$ or $\downarrow$) of a cooper pair, while
the other partner remains almost  immune to the momentum and
energy transfer. However, since a particle's state is a
superposition of two quasiparticle states of both the spins, both
the spin states will be affected in the quasiparticle framework .
The presence of the $\delta$-function in the integrand reveals
that $S({\mathbf q},\delta)$ would be nonzero only when $\hbar
\delta
> 2\Delta$. The integration should be carried out subject to the
restrictions $k \le k_F$ and $k' > k_F$ resulting from Pauli
blocking. Following the method of Ref\cite{abrikosov}, we evaluate
the integral. After a lengthy algebra (the method of calculation
is described in the appendix),  one can express the Eq.
(\ref{dsfu}) in the form \bearr S({\mathbf q},\delta) &=&
\frac{\nu_F}{2}\frac{\Delta^2}{p_qv_F\hbar \delta}
\nonumber \\
 &\times& \int_0^{z_0} dz \frac{(1 +
 j|z|)^2}{(1-jz^2)^{3/2}(1-z^2)^{1/2}}, \label{sint}
\eearr where $\nu_F$ is the density of states at the Fermi
surface. $p_q = \hbar q$, $j = 1- 4\Delta^2/(\hbar\delta)^2$ and
\bearr z_0 = {\rm Min}\left [1, \frac{p_q v_F}{(\hbar^2 \delta^2 -
4\Delta^2)^{1/2}}\right ].  \eearr If $2\Delta < \hbar\delta <
(p_qv_F)^2 + 4\Delta^2)^{1/2}$, then $z_0 = 1$ and the result is
\bearr &S({\mathbf q},\delta)& =
\frac{\nu_F}{8}\frac{\delta}{qv_F}
\nonumber \\
&\times& \left [E(j) + \frac{j}{4}\{8 +
M\hspace{0.05cm}_2F_1(3/2,3/2;2,j)\} \right ] \label{z1} \eearr
where $M = \pi j (1-j)$. Here $E(j)$ represents the complete
elliptic integral and $_2F_1(a,b;c,d)$ is the hypergeometric
function. In the limit $\Delta \rightarrow 0$, $S(\delta, {\mathbf
q}) = \nu_F\delta/(2qv_F)$ which is exactly the dynamic structure
function of {\it normal fluid} \cite{nozpines}. For $\hbar \delta
> (p_qv_F)^2 + 4\Delta^2)^{1/2}$, $S({\mathbf q},\delta)$ can be expressed in terms of
the Elliptic integrals of first and second kind. In this case, as
$\Delta \rightarrow 0$, $S({\mathbf q},\delta) \rightarrow 0$
 and becomes
independent of $q$.  Since this case is not suitable for obtaining
information about the gap, we henceforth focus our attention only
on the former case, that is, $2\Delta < \hbar \delta < (p_qv_F)^2
+ 4\Delta^2)^{1/2}$.

\section{ Dynamic structure function of a trapped superfluid}
Let us now turn our attention to the dynamic structure function of
 a superfluid trapped Fermi gas. This can explicitly be written as
\bearr S({\mathbf q},\delta) &=& \sum_{n,m} \mid \int d^3{\mathbf
x} u_n^*({\mathbf x})v_{m}^*({\mathbf x})\exp(i{\mathbf
q}.{\mathbf x})\mid^2 \nonumber
\\
&\times& \delta(\delta - E_n - E_m) \label{dsfnu} \eearr  In  the
case  of a uniform gas, we have $u_{\mathbf k}({\mathbf x}) =
u_k\exp(i{\mathbf k}.{\mathbf x})$ and $v_{-\mathbf k}({\mathbf
r}) = v_k\exp(-i{\mathbf k}.{\mathbf x})$, and hence the dynamic
structure function defined by Eq. (\ref{dsfnu}) takes the form of
Eq. (\ref{dsfu}). In the LDA, it can be expressed as \bearr
S_{LDA}({\mathbf q},\delta) &=& \frac{1}{(2\pi)^3}\int d^3{\mathbf
x}\int d^3{\mathbf k} n_{k'}({\mathbf x})[1-n_{k}({\mathbf x})]
\nonumber
\\
&\times& \delta(\delta - E_{k'}({\mathbf x}) - E_{k}({\mathbf x}))
\eearr where $n_{k}({\mathbf x})$ is the local momentum
distribution defined by \bearr n_{k}({\mathbf x}) = |v_k({\mathbf
x})|^2 = \frac{1}{2}[1 - \xi_k({\mathbf x})/E_k({\mathbf x})].
\eearr Here $\xi_k({\mathbf x})= \hbar^2k^2/(2m) - \mu({\mathbf
x})$, $\mu({\mathbf x})= \mu - V_{ho}({\mathbf x})$ and
$E_k({\mathbf x}) = \sqrt{\xi_k({\mathbf x})^2 + \Delta({\mathbf
x})}$.  The momentum distribution of trapped Fermi atoms has been
already calculated in Ref\cite{pitaevskii}. After performing the
integration over $k$ as in the preceding section, one can  obtain
\bearr &S_{LDA}&({\mathbf q},\delta) = \frac{1}{2}\int d^3{\mathbf
x} \bar{n}({\mathbf x}) \frac{\Delta({\mathbf
x})^2}{p_qv_F({\mathbf x})\hbar \delta}
\nonumber \\
 &\times& \left [E(j_{{\mathbf x}}) + \frac{j_{{\mathbf x}}}{4}\{8 +
M\hspace{0.05cm}_2F_1(\frac{3}{2},\frac{3}{2};2,j_{{\mathbf x}})\}
\right ] \eearr where $\bar{n}({\mathbf x}) = (3/2)n({\mathbf
x})/\epsilon_F({\mathbf x})$ is the local density of states per
unit volume. Here $j({\mathbf x}) = 1- 4\Delta({\mathbf
x})^2/(\hbar\delta)^2$. The integration over the volume must be
carried out subject to the boundary condition $2\Delta({\mathbf
x}) < \hbar\delta < [(p_qv_F({\mathbf x}))^2+(2\Delta({\mathbf
x}))^2]^{1/2}$. This means that, $\int d^3{\mathbf x} \equiv
2\pi\int_{r_{min}}^{r_{max}}r dr
\int_0^{R_z\sqrt{1-r^2/R_{\perp}}} dz $ where $r_{min}=0$ if
$2\Delta(r=0,z=0) < \hbar\delta $, otherwise it is the solution of
the equation $2\Delta(r,z=0) = \hbar\delta $. Here
$r_{max}<R_{\perp}$ is the solution of the equation
$[(p_qv_F(r,z=0))^2+(2\Delta(r,z=0))^2]^{1/2} = \hbar\delta$.

In the BCS limit $(k_F a_s \rightarrow 0^{-}$), the gap is
exponentially small and can be expressed by the well known formula
\be \Delta_{\rm BCS} \simeq \frac{8
\epsilon_F}{e^2}\exp(-\frac{\pi}{2k_F |a_s|}). \label{bcs} \ee In
calculating $S(\delta,{\mathbf q})$ of trapped atoms in the BCS
limit, one can follow the same method of calculations as described
above. However, the parameter  $k_F |a_s|$ in the exponent of Eq.
(\ref{bcs}) may be approximated by  an average  value.

\section{results and discussions}
Figure 1 shows the results of our calculations. Plotted is the
dynamic structure function $S(\delta,{\mathbf q})$ of superfluid
trapped Fermi atoms as a function of energy transfer for different
values of momentum transfer $q$. $S(\delta,{\mathbf q})$ has been
scaled by  $\nu_F^0 = (3/2)V_{TF}(k_F^0)^3/(6\pi^2\epsilon_F^0)$,
where $V_{TF} = \pi R_{\perp}^2R_{z}$. In the unitarity limit, the
behavior of $S(\delta,{\mathbf q})$ is quite different from that
of superfluid in the weak-coupling BCS limit. This can be
attributed to the occurrence of large gap in the unitarity limit.
In plotting dashed-dotted curve corresponding to the BCS limit, we
have taken $k_F|a_s|= 0.3$, i.e., $\Delta_{\rm BCS} \simeq 0.006
\epsilon_F$. For this small value of the gap, $S(\delta,{\mathbf
q})$ reduces to almost that of normal fluid. We have checked this
by calculating $S(\delta,{\mathbf q})$ for a trapped normal fluid;
and $S(\delta,{\mathbf q})$ for normal fluid is almost
indistinguishable from that of BCS superfluid. This can also be
checked, as discussed in Sec.III, by taking the limit $\Delta
\rightarrow 0$ for which $S(\delta, {\mathbf q})$ reduces to that
of normal fluid.

\begin{figure}
 \includegraphics[width=3.25in]{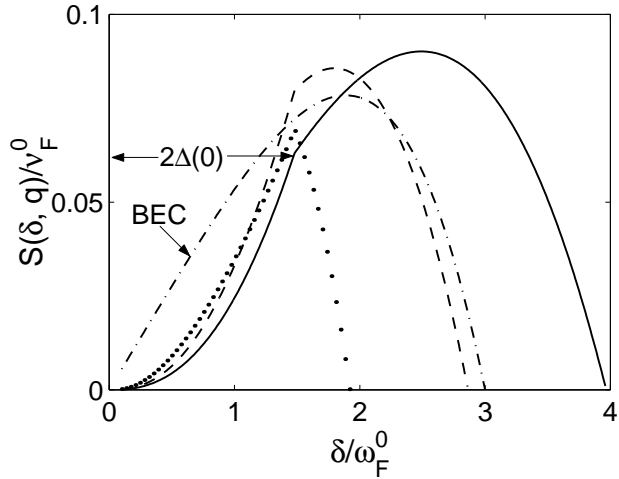}
 \caption{Dimensionless dynamic structure function $S(\delta,{\mathbf
 q})/\nu_F^0$ of a trapped superfluid Fermi gas
 is plotted as a function of dimensionless energy
 transfer $\delta/\omega_F^0$ for the momentum transfer $q/k_F = 1.5$
 (solid), $q/k_F=1$ (dashed), $q/k_F=0.5$ (dotted) in the
 unitarity limit with $\beta = -0.56$ and $\Delta =
 0.49\epsilon_F = 0.74\epsilon_F^0$. For a comparison, $S(\delta,{\mathbf
 q})/\nu_F^0$ in the BCS limit (dashed-dotted) with $|a_s|k_F=0.3$ is
 plotted for $q/k_F = 1.5$. The shift for the unitarity limit (solid) compared to the BCS
 one (dashed-dotted) is $0.8\Delta$ }
 \label{figdel}
 \end{figure}

 Particularly distinguishing feature of $S(\delta,{\mathbf q})$ in
 the unitarity-limited strongly interacting regime as
 compared to normal or BCS regime is the shift of the peak. This
 can be noticed by comparing the dashed-dotted (BCS) and solid
 (unitarity)
 curves which are plotted for the same momentum transfer
 ($q/k_F^0=1.5$). This shift is proportional to the gap.
 For the parameters chosen in Fig.1, the shift is $\sim
 0.8\Delta = 0.8\times 0.49 \epsilon_F = 0.8\times 0.49\times
 (1+\beta)^{-1/2}\epsilon_F^0 \simeq 0.6 \epsilon_F^0$. It should
be mentioned here that the dynamic structure function of a
superfluid trapped Fermi gas in the BCS limit has been evaluated
earlier using a different method of calculation \cite{castin}
which also shows no conspicuous shift except the appearance of an
asymmetric peak.

Another interesting feature is the
 discontinuity in
 $S(\delta,{\mathbf q})$ at energy $\hbar \delta = 2\Delta(0)$,
 where $\Delta(0)$ refers to the gap at the trap center. For a uniform
 Fermi superfluid, $S(\delta,{\mathbf
 q})$ remains zero until energy transfer exceeds $2\Delta$ at
 which it rises sharply
with the increasing energy transfer. For a superfluid trapped
Fermi gas, owing to the spatial distribution of the gap, the
dynamic structure function has a structure below $2\Delta(0)$. As
the energy transfer goes to  zero, the gradient of
$S(\delta,{\mathbf q})$ vanishes. In the low energy regime
($\hbar\delta <2\Delta(0)$), $S(\delta,{\mathbf q})$ varies with
energy nonlinearly. When the energy transfer approaches to
$2\Delta(0)$, the gradient changes abruptly implying the
discontinuity. This behavior can be explained by considering the
boundary conditions $2\Delta({\mathbf x}) < \hbar\delta <
[(p_qv_F({\mathbf x}))^2+(2\Delta({\mathbf x}))^2]^{1/2}$. The
lower bound on $\hbar\delta$ implies that, when $\hbar\delta$ is
less than $2\Delta(0)$, the atoms at the central region of the
trap can not respond to the Bragg pulses, only those atoms in the
peripheral region can be responsive. As $\hbar\delta$ increases
from the origin, increasing number of atoms from the inner region
of the trap start responding. As $\hbar\delta \rightarrow
2\Delta(0)$, the atoms at the trap center can respond. On the
other hand, due to upper bound on $\hbar\delta$,  as $\hbar\delta$
increases, increasing number of atoms from the peripheral region
will cease to respond. Thus, as $\hbar\delta$ exceeds
$2\Delta(0)$, for large momentum transfer,  $S(\delta,{\mathbf
q})$ will rise to a maximum and then eventually vanish  as
$\hbar\delta$ reaches its upper bound
$[(p_qv_F(0))^2+(2\Delta(0))^2]^{1/2}$.

In the recent experiments \cite{kmo,expt2,expt3} with
two-component $^6$Li atoms, the typical value of the Fermi
velocity is $v_F = \hbar k_F/m \sim$ 15 cm/second. The wavelength
for transition 2S$_{1/2}\rightarrow$ 2P$_{3/2}$ in Li atoms is
 $\lambda \sim  670.8$ nm. With counter-propagating Bragg pulses
 tuned near this transition, the momentum transfer would be
${\mathbf q} \simeq 2 k_L \hat{{\mathbf x}}$, where $k_L =
2\pi/\lambda$. This momentum transfer raises the velocity of the
scattered atoms by $2 \times \hbar k_L/m \simeq 20 $ cm/sec which
exceeds $v_F$. Therefore, the scattered atoms should be
distinguishable in time of flight images and hence dynamic
structure function should be measurable. The
polarization-selective Bragg spectroscopy as discussed in Sec.II
may lead to better precision in time-of-flight spin-selective
measurements \cite{molecules,colorado} of the scattered atoms.
Because, the scattered atoms at low temperature would be almost
noninteracting.

Although, we have not studied dynamic structure function in the
BEC limit ($a_sk_F \rightarrow 0^+$) in which Fermi atoms form
bosonic molecules which then condense to form a molecular BEC, the
basic features of the dynamic structure function in this limit can
be understood intuitively. Unless the energy transfer exceeds the
molecular binding energy $\epsilon_m = \hbar^2/(ma_s^2)$, the
molecules would not be dissociated into the atoms. For energy
transfer greater than the binding energy, the dissociated atoms
would have normal Fermi distribution, because they are repulsively
interacting ($a_s$ is positive). Hence, dynamic structure function
would be similar to that of normal fluid except the small shift
$\epsilon_m$ which may not be even discernible.

\section{Conclusion}

In conclusion, we have studied the Bragg scattering of light in a
superfluid trapped Fermi gas in the unitarity limit. Our results
suggest that it is possible to detect the pairing gap in this
limit by large-angle (i.e., large $q$) Bragg scattering. At small
momentum transfer, the scattered atoms may not be distinguishable
in the time-of-flight images. Hence, it  may be difficult to
observe experimentally all the features of the dynamic structure
function in the low energy regime.  In the weak-coupling BCS
limit, the gap is so small that it would elude detection. The
possibility of detecting pairing gap in the unitarity limit has
been earlier discussed qualitatively in Ref \cite{mishra}. In this
paper, we have provided quantitative justification of this
possibility. More exact calculations should involve solving
Bogoliubov-de-Gennes equations and the gap equation at finite
temperature for the entire range of interaction strength.

\acknowledgments{The author is grateful to G. S. Agarwal  and S.
Dutta Gupta for encouragement and discussions, and to M. Randeria
for a helpful suggestion. He is thankful to P. K. Panigrahi, H.
Mishra, A. Mishra and A. Dilip K. Singh for many stimulating
discussions.}

\appendix
\section{}

We here outline the method of calculation of the integral in Eq.
(\ref{dsfu}).
  The dominant contribution to the integration
comes from the $k$-values near the Fermi surface. Therefore,
restricting the integration near $\xi = \epsilon_k -\mu \simeq 0$,
the Eq. (\ref{dsfu})  can be reexpressed as \bearr S({\mathbf
q},\delta) &=& \frac{V}{4(2\pi)^3}\int d^3{\mathbf k} \int
d\xi \delta(\xi_k - \xi) F({\mathbf q},\xi) \nonumber \\
&\simeq& \frac{V}{4(2\pi)^3}\int d^3{\mathbf k}\delta(\xi_k )
\int_{-\infty}^{\infty} d\xi
 F({\mathbf k},\xi)
 \eearr where $ F({\mathbf k},\xi) = \frac{(E' + \xi')(E
- \xi)}{E'E}$ with $E = (\xi^2 + \Delta^2)^{1/2}$ and $E' =
(\xi'^2 + \Delta^2)^{1/2}$. It is of advantage to change  the
variable of integration into $E$ by using the relation $d\xi =
EdE/(E^2 - \Delta^2)^{1/2}$. Considering $E'$ as a function of $E$
and defining the function ${\cal F}(E) = E + E'(E)$, one can use
the identity \bearr \delta(\omega - E-E') =
\frac{\delta(E-E_0)}{\mid d{\cal F}/dE \mid_{E=E_0}} \eearr where
$E_0$ is the solution of the equation ${\cal F}(E) = 0$. After a
lengthy algebra as in Ref\cite{abrikosov}, one finally obtains
\bearr S({\mathbf q},\delta) &=& \frac{V}{(2\pi)^3}
\int d^3{\mathbf k} \delta(\xi_k) \nonumber \\
&\times&\left [\frac{\Delta^2 (\hbar \delta + |{\mathbf
p}_q.{\mathbf v}_k|)^2}{{\cal D}^{3/2}{\cal D'}^{1/2}}\right]
\eearr where ${\mathbf p}_q = \hbar {\mathbf q}$, $v_k = \hbar
k/m$,  ${\cal D} = \hbar^2 \delta^2 - ({\mathbf p}_q.{\mathbf
v}_k)^2$ and ${\cal D'} = {\cal D} - 4\Delta^2$. Writing
$d^3{\mathbf k} = (2 \pi m^2/\hbar^3) d\epsilon_k v_k \sin \theta
d\theta$, where $\theta$ being the angle between ${\mathbf k}$ and
${\mathbf q}$, the integration over $\epsilon_k$ results in \bearr
S({\mathbf q},\delta) =\frac{1}{2} \nu_F \Delta^2 \int dx
\frac{(\hbar \delta + p_qv_F|x|)^2}{D_{F}^{3/2}\sqrt{{\cal D'}_F}}
\label{sx} \eearr where $x=\cos \theta$ and $\nu_F =
(3N_{\sigma}/2\epsilon_F)$ is the density of states of the single
spin $\sigma$. Here subscript $F$ implies that the functions
 ${\cal D}$ and ${\cal D}'$ are evaluated at the
Fermi surface, that is, at $v_k = v_F$. Changing the variable of
integration into \bearr z = \frac{p_q v_F x}{(\hbar^2 \delta^2 -
4\Delta^2)^{1/2}}, \eearr one can express the Eq. (\ref{sx}) in
the form of Eq. (\ref{sint})

\end{document}